\documentclass{iau} 
\usepackage{graphicx}

\title[Pulsating stars in the inner Galaxy] 
{Large-Scale Surveys of \\ Pulsating Stars for Studying \\ Stellar Populations in the Inner Galaxy}

\author[Noriyuki Matsunaga]   
{Noriyuki Matsunaga$^1$}
\affiliation{$^1$Department of Astronomy, The University of Tokyo, 7-3-1 Hongo, Bunkyo-ku, Tokyo 113-0033, Japan \\
email: {\tt matsunaga@astron.s.u-tokyo.ac.jp}} 

\pubyear{2017}
\volume{334}  
\jname{Rediscovering our Galaxy}
\editors{C. Chiappini, I. Minchev, E. Starkenburg, \& M. Valentini, eds.}
\begin{document}

\maketitle

\begin{abstract}
Surveys of pulsating stars in the inner Galaxy have been very limited,
but recent large-scale surveys are rapidly bringing
us new samples of various kinds of variable stars and
new insights into stellar populations therein.
Because of the severe interstellar extinction along the Galactic disc,
the stellar populations in the inner Galaxy are more easily observed
in the infrared,
but even in the infrared the interstellar extinction may cause
a serious problem in revealing their accurate characteristics.
Here we review recent discoveries of Cepheids and Miras,
two kinds of luminous pulsating stars with period--luminosity relation,
in the inner Galaxy.
\keywords{
stars: AGB and post-AGB,
stars: carbon,
stars: distances,
(stars: variables:) Cepheids,
stars: variables: other,
(ISM:) dust, extinction,
Galaxy: bulge,
Galaxy: disc,
Galaxy: stellar content,
Galaxy: structure
}
\end{abstract}

\firstsection 
\section{Introduction}

Classical pulsating stars such as Cepheids and Miras are
useful tracers of the Galaxy. Period-luminosity relations of
those objects allow us to determine the distances to individual stars
(see \cite[Matsunaga 2017]{Matsunaga-2017b}, and references therein).
Each group of pulsating stars has a typical range of ages,
so that we can learn ages of stellar populations in galaxies
based on pulsating stars. Classical Cepheids are young,
Miras are intermediate to old, and type II Cepheids and RR Lyrs are
all old. Especially, ages of classical Cepheids are anti-correlated
with pulsation periods, and that makes it possible to estimate
their accurate ages (\cite[Bono \etal\ 2005]{Bono-2005}).
Furthermore, when appropriate datasets available, we can combine
kinetic and chemical parameters of these objects to reconstruct
chemodynamical information for tracing the structure
and evolution of the Galaxy.

There are some complications which should be borne in mind when
pulsating stars are used as tracers of the Galaxy.
Firstly, classification of the variability type is not always easy
and should be based on good data.
For example, it is possible to get confused between pulsating stars,
eclipsing binaries, and objects with bright or dim stellar spots;
see, e.g., a controversy about faint objects towards the Galactic disc found in
\cite[Chakrabarti \etal\ (2015, 2017)]{Chakrabarti-2015,Chakrabarti-2017}
and \cite{Pietrukowicz-2015b}.
Secondly, stellar rotation may affect ages of Cepheids
and maybe other stars as discussed by
\cite[Anderson \etal\ (2014, 2016)]{Anderson-2014,Anderson-2016}.
Thirdly, evolution in binary systems complicates the implication of
pulsating stars concerning stellar populations. The ages of such objects
can be totally different from those of isolated stars but with
similar variable characteristics. For example,
\cite{Feast-2013} found a carbon-rich Mira in an old-aged
globular clusters. Such old stars are not expected to evolve into
carbon stars but a merged star can become a carbon star. Such objects
must be relatively small in number, but when we discuss rare objects
we need to take such evolutionary paths into consideration.
We'll come back to this point later.
Another interesting case of unusual pulsating stars evolved in a binary system
is found in \cite{Pietrzynski-2012}.
While these complications require further investigations
to address their impacts and necessary corrections, if possible,
on studies on the Galaxy, at least a large fraction of pulsating stars 
are useful for studying stellar populations and their distributions
in the Galaxy.

While pulsating stars have been actively used for studying the Galactic halo
(e.g.~\cite[Huxor \& Grevel 2015]{Huxor-2015};
\cite[Fiorentino \etal\ 2017]{Fiorentino-2017}),
our focus in this review is on the inner part of the Galaxy. 
The use of pulsating stars as tracers of the inner part
and the disc of the Galaxy has been
limited by the incompleteness of surveys (e.g.\  see figure~1 of
\cite[Matsunaga 2012]{Matsunaga-2012}).  
The main reasons for the incompleteness are twofold:
(1) early surveys were optical and relatively shallow
and thus were largely affected by the interstellar extinction
in the disc region,
and (2) sky coverage of sufficiently deep recent surveys, in particular
those in the infrared, has been limited.
In terms of the sky coverage, \textit{Gaia} will produce
an unprecedented catalogue of pulsating stars in the entire range of
the Galaxy, although the interstellar extinction limits its reach across
the Galactic plane. \cite{Windmark-2011} predicted
that the total number of Cepheids in the Galaxy is roughly 20,000 based on
a simple exponential-disc model and about half of them
would be detected by \textit{Gaia}.
There will be many regions
where Cepheids are too much obscured for \textit{Gaia};
in such obscured regions, one cannot expect to get stellar distribution
based on \textit{Gaia}'s parallaxes, and pulsating
stars found in the infrared will be crucial even in
the \textit{Gaia} era. 

\begin{figure}[hbtp]
\begin{center}
\includegraphics[width=0.98\hsize]{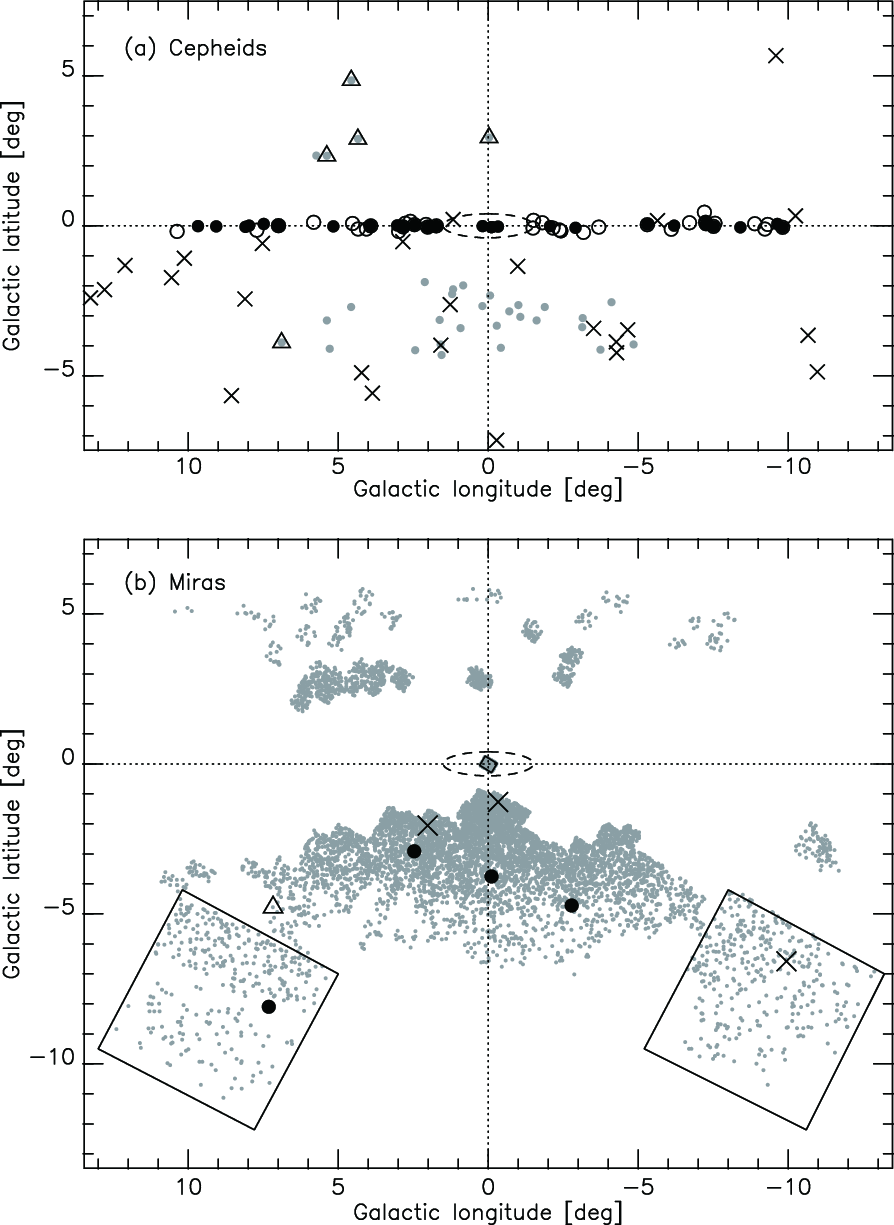} 
\caption{
Distributions of known Cepheids (panel a) and Miras (panel b).
The dashed ellipse around $(l,b)=(0,0)$ suggests the approximate
range of the Nuclear Stellar Disk
(\cite[Launhardt \etal\ 2002]{Launhardt-2002}).
The objects included in each panel are as follows:
(a)~Grey dots indicate classical Cepheids from OGLE
(\cite[Soszy\'nski \etal\ 2011]{Soszynski-2011}),
among which five indicated by triangles are discussed in \cite{Feast-2014a}.
Open and filled circles indicate those found by
\cite[D\'ek\'any \etal\ (2015ab)]{Dekany-2015a,Dekany-2015b}
and by \cite{Matsunaga-2016}, respectively.
Within the dashed ellipse for the NSD located are four Cepheids discussed in
\cite[Matsunaga \etal\ (2011, 2015)]{Matsunaga-2011,Matsunaga-2015}.
Cepheids in this region in the catalogue of \cite{Fernie-1995}
are indicated by crosses, but they are all located within 3~kpc of the Sun
and are foreground objects.
(b)~Grey dots indicate Miras from OGLE
(\cite[Soszy\'nski \etal\ 2013]{Soszynski-2013})
and those discussed in \cite{Catchpole-2016} whose main targets are
located within the two rectangles.
Four filled circles indicate the carbon-rich Miras reported in
\cite{Matsunaga-2017a}, while the cross and triangle symbols indicate
those which are probably foreground and background of the bulge
as discussed in the same paper.
}
\label{fig:LB}
\end{center}
\end{figure}

In this review, we summarize the currently known samples
of Cepheids and Miras in the inner Galaxy
in the sections \ref{sec:Cep} and \ref{sec:Mira}, respectively.
Fig.\,\ref{fig:LB} summarizes currently known Cepheids and Miras
towards the Galactic bulge.
Unless otherwise mentioned, our discussions are limited to this range
of the sky. At the distance to the Galactic centre,
{$\sim$}8.3~kpc (\cite[de Grijs \& Bono 2016]{deGrijs-2016}),
the horizontal stretch of Fig.\,\ref{fig:LB}, 27~degrees, corresponds to
approximately 4~kpc. The dominant component in this range, especially
in terms of stellar mass ($2\times 10^{10}$~M$_\odot$,
\cite[Valenti \etal\ 2016]{Valenti-2016}),
is the bulge with the bar structure inclined by
25--30~degrees to the direction of the Sun
(\cite[Bland-Hawthorn \& Gerhard 2016]{BlandHawthorn-2016}). 
However, several systems with different characteristics are overlapped
in this range, which sometimes complicates interpretation of
observation data on objects in this direction.  
At the very centre exists the Nuclear Stellar Cluster surrounding
the supermassive blackhole, and this cluster is within 
the Nuclear Stellar Disc (or also known as the Central Molecular Zone).
The former has ${\sim}2\times 10^7$~M$_\odot$ in stars
within the radius of 10~pc,
and the latter has ${\sim}10^9$~M$_\odot$, within the radius of 200~pc,
of which a few percent are found in interstellar gas and dust
(\cite[Launhardt, Zylka, \& Mezger]{Launhardt-2002}).
The size of the NSD is illustrated by the ellipse 
in Fig.\,\ref{fig:LB}.  
In the foreground and background extended is the Galactic disc.
These different systems host different stellar populations
and thus different sets of pulsating stars as we see below.

RR Lyrs and type II Cepheids are useful tracers of old stellar population
and have been found in the bulge, although they are outside the scope of
this review. 
Readers are referred to recent observational results in the following papers:
\cite[Pietrukowicz \etal\ (2012,2015)]{Pietrukowicz-2012,Pietrukowicz-2015a},
\cite{Dekany-2013}, \cite[Gran \etal\ (2015, 2016)]{Gran-2015,Gran-2016},
\cite{Minniti-2016}, and \cite{Dong-2017} concerning RR Lyrs, and 
\cite{Soszynski-2011}, \cite{Matsunaga-2013a}, and \cite{Bhardwaj-2017}
concerning type II Cepheids.
The surveys are rather incomplete for these important types
of pulsating stars (see, e.g.\  figure~2
in \cite[Pietrukowicz \etal\ 2015]{Pietrukowicz-2015}
and figure~6 in \cite[Soszy\'nski \etal\ 2011]{Soszynski-2011}).

\section{Cepheids towards the inner Galaxy}
\label{sec:Cep}

Classical Cepheids are pulsating supergiants with pulsation periods
ranging from one day to {$\sim$}80~days or even longer.
They are 10--300~Myr old and their initial masses are 4--10~M$_\odot$
(\cite[Bono \etal\ 2005]{Bono-2005}).
Their period--luminosity relation is one of the most
important steps of the cosmic distance ladder
(\cite[Freedman \etal\ 2001; Riess \etal\ 2016]{Freedman-2001,Riess-2016}).
Since the pioneering work by Baade
(summarized in \cite[Baade 1956]{Baade-1956}), classical Cepheids are
distinguished from type II Cepheids which are evolved low-mass stars
aged {$\sim$}10~Gyr. We'll consider only classical Cepheids hereinafter.

Classical Cepheids, as representatives of young stars, have been used 
for studying various characteristics of the Galactic disc (e.g.~\cite[Feast \& Whitelock 1997; Majaess, Turner, \& Lane 2009]{Feast-1997,Majaess-2009}).
In particular, Cepheids have been the most successful tracers of
the metallicity gradient, i.e.~how the metallicity changes as
a function of the distance from the Galactic centre
(\cite[Genovali \etal\ 2014, 2015; da Silva \etal\ 2016]{Genovali-2014,Genovali-2015,daSilva-2016}).
However, the previous survey of classical Cepheids is incomplete as we
discussed above. Fig.\,\ref{fig:LB} plots about a dozen of previously
known Cepheids in the DDO database (\cite[Fernie \etal\ 1995]{Fernie-1995}),
but they are all within 3~kpc of the Sun. It is considered that
the bulge is dominated by old stellar populations even if a small fraction
of young stars exist (\cite[Zoccali \etal\ 2003; Clarkson \etal\ 2011; Benzby \etal\ 2013]{Zoccali-2003,Clarkson-2011,Bensby-2013}),
and therefore classical Cepheids are not expected
and haven't been found in the bulge. In contrast, the NSD
is known to host stars with a wide range of ages
(\cite[Serabyn, \& Morris 1996; Figer \etal\ 2004]{Serabyn-1996,Figer-2004}).
\cite[Matsunaga \etal\ (2011, 2015)]{Matsunaga-2011,Matsunaga-2015}
in fact found four classical Cepheids in this system. 
They are located within 0.35~degrees of the Galactic centre,
or within 50~pc (projected distance), and their radial velocities
are consistent with the rotation of the NSD
(\cite[Matsunaga \etal\ 2015]{Matsunaga-2015}).
Interestingly, all of them have periods around 20~days (18.8--23.5~days),
which is significantly longer than the median period, {$\sim$}5~days,
of Galactic Cepheids, and thus are similarly aged {$\sim$}25~Myr,
which puts an important constraint on the star formation history
in this system (\cite[Matsunaga \etal\ 2011]{Matsunaga-2011}).
The foreground interstellar extinction for these Cepheids are
{$\sim$}2.5~mag or more at around 2~{$\mu$}m, corresponding to 
{$\sim$}30~mag in the optical, so that infrared observations 
are required for both surveys and follow-up studies. 

Later, infrared surveys by
\cite[Dekany \etal\ (2015ab)]{Dekany-2015a,Dekany-2015b}
and \cite{Matsunaga-2016} found dozens of Cepheids in the mid plane of
the Galactic disc, i.e.~nearly 0~degree in Galactic latitude
(Fig.\,\ref{fig:LB}).
While \cite{Dekany-2015b} suggested the presence of an inner thin disc of
young stars surrounding the NSD, \cite{Matsunaga-2016}
reported the lack of young stars in the inner 2.5~kpc except the NSD.
As discussed in \cite{Matsunaga-2016}, the distances to these reddened
Cepheids ($E_{H-K_{\rm s}}=$1.2--2.5~mag) have large errors
due to the uncertainty in the extinction law, i.e.~the wavelength dependency
of the interstellar extinction.
The four classical Cepheids in the NSD give an important
constraint on the extinction law because they should be located
at the distance of the Galactic centre regardless of the extinction.
This supports the extinction law of \cite{Nishiyama-2006}
rather than that of \cite{Nishiyama-2009} which was used by 
\cite[Dekany \etal\ (2015ab)]{Dekany-2015a,Dekany-2015b}.
More detailed discussions on the problem of the extinction law
are found in the original paper of \cite{Matsunaga-2016} and
a recent review (\cite[Matsunaga 2017]{Matsunaga-2017b}).
The lack of Cepheids and young stars in the innermost part of
the Galactic disc is supported by the distribution of
observed H~II regions and massive star-forming regions
(e.g.~\cite[Jones \etal\ 2013; Sanna \etal\ 2014]{Jones-2013,Sanna-2014}).
This suggests that there is no simple exponential disc extending
into the centre in contrast to the model used by \cite{Windmark-2011},
if we consider young stars, and we need further infrared surveys to
map this region including the entire NSD and the interface between
the discs and the bulge.

Another interesting sample of classical Cepheids towards
the region of our interest but a few degrees away from the Galactic mid plane
came from a large-scale optical survey,
Optical Gravitational Lensing Experiment (OGLE).
\cite{Soszynski-2011} discovered 32 classical Cepheids towards the bulge
(Fig.\,\ref{fig:LB}) besides hundreds of type II Cepheids. 
Although most of the type II Cepheids belong to the bulge
considering their positions on the period--Wesenheit diagram
(figure~7 in \cite[Soszy\'nski \etal\ 2011]{Soszynski-2011}),
most of the classical Cepheids seem to be located further than the bulge
(their magnitudes are similar to those of type II Cepheids in spite of
the large difference between intrinsic magnitudes for these two types).
Five of the 32 classical Cepheids,
triangles in the panel (a) of Fig.\,\ref{fig:LB}, were investigated by
\cite{Feast-2014a} in more detail, and their distances are
larger than 20~kpc, indeed further than the bulge. Moreover, they are
separated from the Galactic plane by {$\sim$}1~kpc or more and
belong to the flared part of the disc outskirts.
Other Cepheids in \cite{Soszynski-2011} are also located
at similarly large distances and exotic objects like these five,
but their nature needs to be revealed by further investigations.

\section{Miras towards the inner Galaxy}
\label{sec:Mira}

Miras are pulsating giants with periods longer than 100~days.
Their initial masses and ages are relatively wide:
1--9~M$_\odot$ corresponding to between {$\sim$}10~Gyr and 30~Myr
(\cite[Iben 1983]{Iben-1983}).
There is an anti-correlation between periods and ages
(\cite[Feast, Whitelock, \& Menzies 2006]{Feast-2006}), 
although its relation is not so tight or established as 
the period--age relation of classical Cepheids.
Nevertheless, a wide range of the ages of Miras is useful
for tracing old to intermediate-age stellar populations.
Miras also have the period--luminosity relation and serves as
distance indicator, although one can find useful relations only
in the infrared or the bolometric magnitude but not in the optical
(\cite[Glass \& Lloyd Evans 1981; Feast \etal\ 1989; Whitelock \etal\ 2008]{Glass-1981,Feast-1989,Whitelock-2008}).
As has been discussed in \cite{Feast-2014b} and \cite{Whitelock-2014},
Miras are very bright in the infrared and can be a good alternative
to Cepheids as distance indicators to galaxies at large distances
for infrared facilities like
\textit{James Webb Space Telescope (JWST)} in the future.

Because of the high luminosities of Miras, surveys of Miras in the bulge,
especially towards low-extinction windows, have been done from the early days
(see the review given by \cite[Catchpole \etal\ 2016]{Catchpole-2016}).
Recently, \cite{Soszynski-2013} found more than 6500 Miras using
the OGLE dataset (Fig.\,\ref{fig:LB}), which doesn't include
the low-latitude region ($|b|<1^\circ$). Such central parts require
infrared surveys even if they are relatively shallow, and in fact
early surveys in the near infrared already detected dozens or hundreds
of Miras (\cite[Glass \etal\ 2001]{Glass-2001}, and references therein)
within the NSD. \cite{Matsunaga-2009} detected over 500 Miras
(with periods determined) and obtained the distance to the Galactic centre,
8.24~kpc, making use of multi-band photometry in $JHK_{\rm s}$ bands
which was necessary to make correction of the interstellar extinction.
This estimate or any other estimate based on standard candles
(i.e.~luminosity based distance indicators) is affected by
the uncertainty in the extinction law described above
(\cite[Matsunaga 2013]{Matsunaga-2013b}). Even with this uncertainty
in the distances, it is clear that Miras show a concentration
towards the Galactic centre (\cite[Matsunaga \etal\ 2009]{Matsunaga-2009};
see also the large-scale density gradient in Fig.\,\ref{fig:LB}).
It is, however, unclear how Miras in the central 100~pc should be
separated into the NSD and the extended bar-like bulge.

An important group of objects related to Miras, with at least some overlaps,
is sources with maser emissions with SiO, H$_2$O, and OH molecules.
These emissions are produced in evolved stars with thick circumstellar shell,
and those objects are often found to be Miras or related pulsating stars
like semi-regulars (e.g.~see the review by \cite[Habing 1996]{Habing-1996}).
In particular, OH/IR stars characterized by OH maser emission and 
high infrared luminosity have played an important role in studying
intermediate-age populations in the inner Galaxy
(\cite[Lindqvist, Habing, \& Winnberg 1992; Wood, Habing, \& McGregor 1998]{Lindqvist-1992,Wood-1998}).
Their kinematics supports that these intermediate-age objects belong
to the NSD.
Later, \cite{Deguchi-2004} detected SiO maser in 180 Miras of
the catalogue published by \cite{Glass-2001}, 
and also found the rotation comparable to that of OH/IR stars.
This seems to suggest that at least a part of the Miras near the centre
belong to the NSD, but the more detailed studies on
distributions of these Miras and maser sources should be done 
with more comprehensive datasets including kinematic information
from both radial velocities and proper motions.

The recent discovery by \cite{Matsunaga-2017a} has revealed
a rare group of objects, carbon-rich Miras, among the Miras in the bulge.
They selected several candidates of carbon-rich Miras,
among the Miras in \cite{Soszynski-2013} and \cite{Catchpole-2016}
based on the $(J-K_{\rm s})$-$([9]-[18])$ diagram in which 
one can distinguish objects with carbon-rich dust shell from
those with oxygen-rich dust shell (\cite[Ishihara \etal\ 2011]{Ishihara-2011}).
Then, with spectroscopic follow-up observations,
they confirmed 8 carbon-rich Miras of which 3 are foreground objects.
These carbon-rich Miras are considered to represent
relatively (or totally) unexplored stellar population(s).
Both age and metallicity affect whether an AGB star is evolved into
carbon-rich stars or not, and intermediate-age stars (a few Gyr)
tend to finish the AGB phase as carbon stars (\cite[Marigo 2013]{Marigo-2013}). 
For the Galactic bulge, no carbon-rich AGB stars had been confirmed before.
This was naturally understood because the dominant population
of the bulge is old and metal-rich. Moreover, many stars are expected
to be oxygen enhanced which makes it even harder to produce carbon-rich stars.
Unfortunately, the ages and the origin of the carbon-rich Miras found in
the bulge are still unclear because mass-augmented stars may
evolve into carbon-rich stars even if they are old
(\cite[Feast \etal\ 2013]{Feast-2013}).
Kinematics and chemical information may help uncover their origins.

\section{Summary}

In the last decade, large-scale surveys like OGLE and
some near-infrared ones have revealed 
large collections of new Cepheids and Miras.
The same is true for RR Lyrs and other kinds of variable stars.
Nevertheless, Fig.\,\ref{fig:LB} clearly shows
that we still miss a large numbers of these objects
particularly in the low-latitude region, in which
every system of the inner Galaxy is overlapped
(i.e.~the Galactic bulge, the Nuclear Stellar Disk,
and the Nuclear Stellar Clusters) together with the Galactic disc
in the foreground and background.
Several large-scale surveys including
OGLE, VISTA Via Lactea (\cite[Minniti \etal\ 2010]{Minniti-2010}), and
\textit{Gaia}
are expected to provide us with more complete samples of pulsating stars
which will be crucial for studying stellar populations in the future.
Various parameters are readily accessible
using the period--luminosity and period--age relations, for example,
and other observables for the pulsating stars, and they will remain to be
useful tracers of stellar populations in the Galaxy.

\acknowledgement{
The author appreciate financial support from the Japan Society for the Promotion of Science (JSPS) through the Grant-in-Aid, No.~26287028.
}

%
%
%

\begin{thebibliography}{}

\bibitem[Anderson \etal\ (2014)]{Anderson-2014}
   {Anderson, R.I., Ekstr\"om, S., Georgy, C., Meynet, G., Mowlavi, N., Eyer, L.} 2014, \textit{A\&A} 564, A100
\bibitem[Anderson \etal\ (2016)]{Anderson-2016}
   {Anderson, R.I., Saio, H., Ekstr\"om, S., Georgy, C., Meynet, G.} 2016, \textit{A\&A} 591, A8
\bibitem[Alonso Garc\'ia \etal\ (2015)]{Alonso-Garcia-2015}
   {Alonso-Garc\'ia, J., D\'ek\'any, I., Catelan, M., Contreras Ramos, R., Gran, F., Amigo, P., Leyton, P., \& Minniti, D.} 2015, \textit{AJ} 149, 99
\bibitem[Baade (1956)]{Baade-1956}
   {Baade, W.} 1956, \textit{PASP} 68, 5
\bibitem[Bensby \etal\ (2013)]{Bensby-2013}
   {Bensby, T., \etal\ } 2013, \textit{A\&A} 549, A147 
\bibitem[Bhardwaj \etal\ (2017)]{Bhardwaj-2017}
   {Bhardwaj, A., \etal\ } 2017, \textit{A\&A}, in press (arXiv:1707.03755)
\bibitem[Bland-Hawthorn \& Gerhard (2016)]{BlandHawthorn-2016}
   {Bland-Hawthorn, J., \& Gerhard, O.} 2016, \textit{ARA\&A} 54, 529
\bibitem[Bono \etal\ (2005)]{Bono-2005}
   {Bono, G., Marconi, M., Cassisi, S., Caputo, F., Gieren, W., \& Pietrzynski, G.} 2005, \textit{ApJ} 621, 966
\bibitem[Catchpole \etal\ (2016)]{Catchpole-2016}
   {Catchpole, R.M., Whitelock, P.A., Feast, M.W., Hughes, S.M.G., Irwin, M., Alard, C.} 2016, \textit{MNRAS} 455, 2216
\bibitem[Chakrabarti \etal\ (2015)]{Chakrabarti-2015}
   {Chakrabarti, S., Saito, R., Quillen, A., Gran, F., Klein, C., \& Blitz, L.} 2015, \textit{ApJ} (Letters) 802, L4
\bibitem[Chakrabarti \etal\ (2017)]{Chakrabarti-2017}
   {Chakrabarti, S., \etal\ } 2017, \textit{ApJ} 844, 159
\bibitem[Clarkson \etal\ (2011)]{Clarkson-2011}
   {Clarkson, W.I., \etal\ } 2011, \textit{ApJ} 735, 37
\bibitem[da Silva \etal\ (2016)]{daSilva-2016}
   {da Silva, R., \etal\ } 2016, \textit{A\&A} 586, A125
\bibitem[de Grijs \& Bono (2016)]{deGrijs-2016}
   {de Grijs, R., \& Bono, G.} 2016, \textit{ApJS} 227, 5
\bibitem[Deguchi \etal\ (2004)]{Deguchi-2004}
   {Deguchi, S., \etal\ } 2004, \textit{PASJ} 56, 261
\bibitem[D\'ek\'any \etal\ (2013)]{Dekany-2013}
   {D\'ek\'any, I., Minniti, D., Catelan, M., Zoccali, M., Saito, R.K., Hempel, M., Gonzalez, O.A.} 2013, \textit{ApJ} (Letters) 776, L19
\bibitem[D\'ek\'any \etal\ (2015a)]{Dekany-2015a}
   {D\'ek\'any, I., \etal\ } 2015a, \textit{ApJ} (Letters) 799, L11
\bibitem[D\'ek\'any \etal\ (2015b)]{Dekany-2015b}
   {D\'ek\'any, I., \etal\ } 2015b, \textit{ApJ} (Letters) 812, L29
\bibitem[Dong \etal\ (2017)]{Dong-2017}
   {Dong, H., \etal} 2017, \textit{MNRAS} 471, 3617
\bibitem[Feast \etal\ (1989)]{Feast-1989}
   {Feast, M.W., Glass, I.S., Whitelock, P.A., \& Catchpole, R.M.} 1989, \textit{MNRAS} 241, 375
\bibitem[Feast, \& Whitelock (2014)]{Feast-2014b}
   {Feast, M.~W., \& Whitelock, P.A.} 2014, \textit{IAUS} 298, 40
\bibitem[Feast \& Whitelock (1997)]{Feast-1997}
   {Feast, M.W., \& Whitelock, P.A.} 1997, \textit{MNRAS} 291, 683
\bibitem[Feast, Whitelock, \& Menzies (2006)]{Feast-2006}
   {Feast, M.W., Whitelock, P.A., \& Menzies, J.W.} 2006, \textit{MNRAS} 369, 791
\bibitem[Feast, Menzies, \& Whitelock (2013)]{Feast-2013}
   {Feast, M.W., Menzies, J.W., \& Whitelock, P.A.} 2013, \textit{MNRAS} 428, L36
\bibitem[Feast \etal\ (2014)]{Feast-2014a}
   {Feast, M.W., Menzies, J.W., Matsunaga, N., \& Whitelock, P.A.} 2014, \textit{Nature} 509, 342
\bibitem[Fernie \etal\ (1995)]{Fernie-1995}
   {Fernie, J.D., Evans, N.R., Beattie, B., \& Seager, S.} 1995, \textit{Information Bulletin on Variable Stars} 4148, 1
\bibitem[Figer \etal\ (2004)]{Figer-2004}
   {Figer, D.F., Rich, R.M., Kim, S.S., Morris, M., \& Serabyn, E.} 2004, \textit{ApJ} 601, 319
\bibitem[Fiorentino \etal\ (2017)]{Fiorentino-2017}
   {Fiorentino, G., \etal\ } 2017, \textit{A\&A} 599, A125
\bibitem[Freedman \etal\ (2001)]{Freedman-2001}
   {Freedman, W.L., \etal\ } 2001, \textit{ApJ} 553, 47
\bibitem[Genovali \etal\ (2014)]{Genovali-2014}
   {Genovali, K., \etal\ } 2014, \textit{A\&A} 566, A37
\bibitem[Genovali \etal\ (2015)]{Genovali-2015}
   {Genovali, K., \etal\ } 2015, \textit{A\&A} 580, A17
\bibitem[Glass \& Lloyd Evans (1981)]{Glass-1981}
   {Glass, I.S., \& Lloyd Evans, T.} 1981, \textit{Nature} 291, 303
\bibitem[Glass \etal\ (2001)]{Glass-2001}
   {Glass, I.S., Matsumoto, S., Carter, B.S., \& Sekiguchi, K.} 2001, \textit{MNRAS} 321, 77
\bibitem[Gran \etal\ (2015)]{Gran-2015}
   {Gran, F., Minniti, D., Saito, R.K., Navarrete, C., D\'ek\'any, I., McDonald, I., Contrares, R.R., \& Catelan, M.} 2015, \textit{A\&A} 575, A114
\bibitem[Gran \etal\ (2016)]{Gran-2016}
   {Gran, F., \etal\ } 2016, \textit{A\&A} 591, A145
\bibitem[Habing (1996)]{Habing-1996}
   {Habing H.J.} 1996, \textit{A\&AR} 7, 97
\bibitem[Huxor \& Grebel (2015)]{Huxor-2015}
   {Huxor, A.P., \& Grebel, E.K.} 2015, \textit{MNRAS} 453, 2653
\bibitem[Iben \& Renzini (1983)]{Iben-1983}
   {Iben, I.Jr., \& Renzini, A.} 1983, \textit{ARA\&A} 21, 271
\bibitem[Ishihara \etal\ (2011)]{Ishihara-2011}
   {Ishihara, D., Kaneda, H., Onaka, T., Ita, Y., Matsuura, M., \& Matsunaga, N.} 2011, \textit{A\&A} 534, A79
\bibitem[Jones \etal\ 2013]{Jones-2013}
   {Jones, C., Dickey, J.M., Dawson, J.R., McClure-Griffiths, N.M., Anderson, L.D., \& Bania, T.M.} 2013, \textit{ApJ} 774, 117
\bibitem[Launhardt, Zylka, \& Mezger (2002)]{Launhardt-2002}
   {Launhardt, R., Zylka, R., \& Mezger, P.G.} 2002, \textit{A\&A} 384, 112
\bibitem[Lindqvist, Habing, \& Winnberg (1992)]{Lindqvist-1992}
   {Lindqvist, M., Habing, H.J., \& Winnberg, A.} 1992, \textit{A\&A} 259, 118
\bibitem[Majaess, Turner, \& Lane (2009)]{Majaess-2009}
   {Majaess, D.J., Turner, D.G., \& Lane, D.J.} 2009, \textit{MNRAS} 398, 263
\bibitem[Marigo (2013)]{Marigo-2013} 
   {Marigo, P., Bressan, A., Nanni, A., Girardi, L., \& Pumo, M.L.} 2013, \textit{MNRAS} 434, 488
\bibitem[Matsunaga \etal\ (2009)]{Matsunaga-2009}
   {Matsunaga, N., Kawadu, T., Nishiyama, S., Nagayama, T., Hatano, H., Tamura, M., Glass, I.~S., \& Nagata, T.} 2009, \textit{MNRAS} 399, 1709
\bibitem[Matsunaga \etal\ (2011)]{Matsunaga-2011}
   {Matsunaga, N., Kawadu, T., Nishiyama, S., et al.} 2011, \textit{Nature} 188, 477
\bibitem[Matsunaga (2012)]{Matsunaga-2012}
   {Matsunaga, N.} 2012, \textit{Journal of Physics: Conference Series} 372, 12026
\bibitem[Matsunaga 2013]{Matsunaga-2013b}
   {Matsunaga, N.} 2013, \textit{IAUS} 289, 109
\bibitem[Matsunaga \etal\ (2013)]{Matsunaga-2013a}
   {Matsunaga, N., Feast, M.~W., Kawadu, T., et al.} 2013, \textit{MNRAS} 429, 385
\bibitem[Matsunaga \etal\ (2015)]{Matsunaga-2015}
   {Matsunaga, N., Fukue, K., Yamamoto, R., et al.} 2015, \textit{ApJ} 799, 46
\bibitem[Matsunaga \etal\ (2016)]{Matsunaga-2016}
   {Matsunaga, N., Feast, M.~W., Bono, G., et al.} 2016, \textit{MNRAS} 462, 414
\bibitem[Matsunaga (2017)]{Matsunaga-2017b}
   {Matsunaga, N.} 2017, \textit{EPJ Web of Conferences} 152, 1007
\bibitem[Matsunaga \etal\ (2017)]{Matsunaga-2017a}
   {Matsunaga, N., Menzies, J.~W., Feast, M.~W., Whitelock, P.~W., Onozato, H., Barway, S., \& Aydi, E.} 2017, \textit{MNRAS} 469, 4949 
\bibitem[Minniti \etal\ 2010]{Minniti-2010}
   {Minniti, D., \etal\ } 2010, \textit{New Astron.} 15, 433
\bibitem[Minniti \etal\ (2016)]{Minniti-2016}
   {Minniti, D., Contreras Ramos, R., Zoccali, M., Rejkuba, M., Gonzalez, O.A., Valenti, E., \& Gran, F.} 2016, \textit{ApJ} (Letters) 830, L14
\bibitem[Nishiyama \etal\ (2006)]{Nishiyama-2006}
   {Nishiyama, S., \etal\ } 2006, \textit{ApJ} 638, 839
\bibitem[Nishiyama \etal\ (2009)]{Nishiyama-2009}
   {Nishiyama, S., Tamura, M., Hatano, H., Kato, D., Tanab\'e, T., Sugitani, K., Nagata, T.} 2009, \textit{ApJ} 696, 1407
\bibitem[Pietrukowicz \etal\ (2012)]{Pietrukowicz-2012}
   {Pietrukowicz, P., \etal\ } 2012, \textit{ApJ} 750, 169
\bibitem[Pietrukowicz \etal\ (2015a)]{Pietrukowicz-2015a}
   {Pietrukowicz, P., \etal\ } 2015a, \textit{ApJ} 811, 113
\bibitem[Pietrukowicz \etal\  (2015b)]{Pietrukowicz-2015b}
   {Pietrukowicz, P., \etal\ } 2015b, \textit{ApJ} (Letters) 813, L40
\bibitem[Pietrzy\'nski \etal\ (2012)]{Pietrzynski-2012}
   {Pietrzy\'nski, G., \etal\ } 2012, \textit{Nature} 484, 75
\bibitem[Riess \etal\ (2016)]{Riess-2016}
   {Riess, A.G., \etal\ } 2016, \textit{ApJ} 826, 56
\bibitem[Sanna \etal\ 2014]{Sanna-2014}
   {Sanna, A., \etal\ } 2014, \textit{ApJ} 781, 108
\bibitem[Serabyn, \& Morris (1996)]{Serabyn-1996}
   {Serabyn, E., \& Morris, M.} 1996, \textit{Nature} 382, 602
\bibitem[Soszy\'nski \etal\ (2011)]{Soszynski-2011}
   {Soszy\'nski, I., \etal\ } 2011, \textit{AcA} 61, 285
\bibitem[Soszy\'nski \etal\ (2013)]{Soszynski-2013}
   {Soszy\'nski, I., \etal\ } 2013, \textit{AcA} 63, 21
\bibitem[Valenti \etal\ (2016)]{Valenti-2016}
   {Valenti, E. \etal\ } 2016, \textit{A\&A} 587, L6
\bibitem[Whitelock, \& Feast (2014)]{Whitelock-2014}
   {Whitelock, P.A., \& Feast, M.~W.} 2014, \textit{EAS Publications Series} 67--68, 263
\bibitem[Whitelock, Feast, \& van Leeuwen (2008)]{Whitelock-2008}
   {Whitelock, P.A., Feast, M.W., \& van Leeuwen, L.} 2008, \textit{MNRAS} 386, 313
\bibitem[Windmark, Lindegren, \& Hobbs (2011)]{Windmark-2011}
   {Windmark, F., Lindegren, L., \& Hobbs, D.} 2011, \textit{A\&A} 530, A76
\bibitem[Wood, Habing, \& McGregor (1998)]{Wood-1998}
   {Wood, P.R., Habing, H.J., \& McGregor, P.J.} 1998, \textit{A\&A} 336, 925
\bibitem[Zoccali \etal\ (2003)]{Zoccali-2003}
   {Zoccali, M. \etal\ } 2003, \textit{A\&A} 399, 931

\end{thebibliography}
\end{document}